\documentclass{ws-mpla}
\usepackage{amsmath,amssymb}

\begin{document}


%
\catchline{}{}{}{}{}
%

\title{Non-singular radiation cosmological models}

  \author{L. Fern\'andez-Jambrina}
 \address{E.T.S.I. Navales\\ Universidad Polit\'ecnica de Madrid \\
 Arco de la Victoria s/n \\ E-28040 Madrid, Spain\\
 lfernandez@etsin.upm.es \\
 http://debin.etsin.upm.es/lfj.htm}

 \author{L.M. Gonz\'alez-Romero}
 \address{Departamento de F\'\i sica Te\'orica II\\
  Facultad de Ciencias F\'\i sicas \\
  Universidad Complutense de Madrid \\
  Avenida Complutense s/n \\ E-28040 Madrid, Spain\\
 mgromero@fis.ucm.es}


\date{\today}
 \maketitle

\begin{abstract}
In this paper we analyse the possibility of constructing
singularity-free inhomogeneous cosmological models with a pure radiation field as
matter content. It is shown that the conditions for regularity are
very easy to implement and therefore there is a huge number of such
spacetimes. 

\keywords{non-singular; cosmology; geodesic completeness; radiation.}
\end{abstract}

\ccode{PACS Nos.: 04.20.Dw, 04.20.Ex, 04.20.Jb.}

\maketitle

\section{Introduction}

Singularity-free cosmological models were quite popular until the
sixties. At that time there were attempts to circumvent initial and
final singularities, for instance, by introducing additional fields,
such as Hoyle and Narlikar's $C$ field. Such models had the drawback of requiring
continuous creation of matter \cite{hoyle}. The interest for that line of research
decayed in the seventies when most of the famous singularity theorems
were published (cfr. for instance \cite{HE}, \cite{beem}). Under very general and physically realistic conditions
these theorems seemed to push cosmological models towards the appearance
of singularities, that could be of the Big Bang kind or not. Although
most quantum fields, such as Yang-Mills, do not follow them, there
was nothing to object to energy conditions. Less criticism would
receive causality conditions, which require the absence of closed causal
curves. Or the generic condition, which roughly states that the
universe is not empty \cite{beem}.

However, in addition to those general and physically acceptable
requirements, there were other conditions that might be overcome.
These were the existence of some causally trapped set in the
spacetime, such as closed trapped surfaces \cite{penrose} or an achronal set without
edge \cite{hawking}. The fulfillment of the latter conditions is not so obvious as
for the former ones.

But it was not until the nineties when the first singularity-free
cosmological models began to appear \cite{seno}. These models were
not realistic, since they were inhomogeneous spacetimes with
cylindrical symmetry, but they pointed out that on diminishing the
symmetry non-singular behaviour might appear. Since these models were
exact solutions in a framework where they are fairly difficult to
obtain, it is not strange that the list of such non-singular models
was rather short until the end of the century \cite{grg}. In fact these
spacetimes belonged to larger families among which the absence of
singularities required a certain amount of fine tuning \cite{esc}.
Consequently it was tempting to consider singularity-free cosmological models as a
curiosity.

However a large family of regular cosmological models was
recently found \cite{wide} and did not depend on a finite number of parameters but on
two nearly arbitrary functions. This result could mean
that although our universe has an initial singularity, it is yet to be
explained why alternative, and abundant, possibilities were not
considered.

The aim of this paper dates back to Senovilla's original solution.
This spacetime had the realistic property of being filled with an
incoherent radiation perfect fluid, which is the standard mean matter
content for a universe in its early epoch. It is intriguing that
when the Senovilla spacetime was extended to include other
models with also a linear equation of state \cite{ruiz}, again radiation fluids
were the only ones which remained non-singular \cite{ruiz}. What we
plan to discuss now is the role of radiation in non-singular behaviour.
But instead of considering incoherent radiation, we shall focus on
pure radiation.

With this aim in mind we devote the next section of this paper to the
Einstein equations for radiation fields in inhomogenous cosmological
models with a two-dimensional group of isometries acting on spacelike
orbits. Section 3 comprises three ways of simplification of the
equations. One of them, corresponding to models with a transitivity
surface element with lightlike gradient, incorporates additional
symmetries to the spacetimes. In Section 4 the analysis of the
possible singularities of the families described in the previous
sections is completed. A discussion of the results is included at the
end of the paper.

\section{Inhomogeneous radiation cosmological models}

We may describe pure radiation by its conserved energy-momentum
tensor,

\begin{equation}
    T^{\mu\nu}=\Phi k^{\mu}k^\nu,
    \label{null}
\end{equation}
where $k$ is a null vector field describing the lightlike 4-velocity
of the radiation, $k^\mu k_{\mu}=0$, and $\Phi$ is a positive function
related, but not equal, to the radiation energy, since this function
could be reabsorbed by the vector field $k$ without disturbing the
properties of the tensor. Bear in mind that
radiation fields are characterised by having all energy-momentum
tensor invariants equal to zero.

The energy-momentum conservation equations may be written,

\begin{equation}
    T^{\mu\nu}_{\ ;\nu}=\Phi_{,\nu}k^\nu k^{\mu}+\Phi a^{\mu}+\Phi
    \Theta k^{\mu}=0,
    \label{conserved}
\end{equation}
in terms of the expansion $\Theta=k^\nu_{;\nu}$ and acceleration,
$a^\mu=k^\mu_{;\nu}k^{\nu}$ of the lightlike congruence. Since the
acceleration is orthogonal to the radiation velocity, $a$ might have
a component parallel to $k$ and a spatial component orthogonal to $k$.
The existence of the latter component would mean that the equations
would split into components parallel and non-parallel to $k$ and
$\Phi$ would be equal to zero. Since this is not our case, $a$ is
proportional to $k$ and (\ref{conserved}) is a single-component
equation.

Now we turn to the geometric description of the spacetime. Considering that
except for some spherically symmetric spacetimes \cite{grg} every
other singularity-free cosmological model has been found within the
Ansatz of an Abelian two-dimensional orthogonally transitive group of isometries acting on
spacelike orbits, we shall impose that our models share these
symmetries. We choose as generators for the symmetries two commuting fields,
that we write as $\{\partial_{y},\partial_{z}\}$, having in mind
the coordinates that we want to use. We take them as mutually
orthogonal. There is much freedom for
choosing the rest of the coordinates, which we shall name $t$, $x$.
Since every surface admits and isothermal parametrization for which
the metric is conformally flat, we take as our $t$ and $x$ the
coordinates in such parametrization. With this choice of coordinates,
the metric can be written in a suitable form \cite{kramer},
\begin{equation}
ds^2=e^{2K}(-dt^2+dx^2)+\rho^2e^{2U}dy^2+e^{-2U}dz^2,\label{metric}
\end{equation}
 depending on three functions, $K$, $U$ and $\rho$, of just $t$ and $x$.

As we shall see, there is no possibility for having an axis in these
models. Therefore we reject the interpretation of the isometries as cylindrical
symmetry and there is no reason a priori for restricting the ranges of the
coordinates,
\begin{equation}
    -\infty<t,x,y,z<\infty.
\end{equation}

As an Ansatz we take a velocity of the radiation with no projection on
the Killing fields, that is,

\begin{equation}
    k_{\pm}=e^{-K}(\partial_{t}\pm\partial_{x}),
\end{equation}
where the plus (minus) sign stands for growing (decreasing)
coordinate $x$.
The normalization factor has been chosen so that $\langle
k_{+},k_{-}\rangle=-2$.

For such velocities the energy-momentum conservation
(\ref{conserved}) is written as

\begin{equation}
\Phi_{t}\pm
\Phi_{x}+\Phi\left\{\frac{\rho_{t}\pm\rho_{x}}{\rho}+2(K_{t}\pm K_{x})\right\}=0,
    \label{cons}
\end{equation}
which allows a rather simple form in terms of ingoing or outgoing
lightlike coordinates,
\begin{equation}
\left(\partial_{t}\pm\partial_{x}\right)\left\{\Phi e^{2K}\rho\right\}=0,
    \label{cons1}
\end{equation}
for which integration is straightforward,
\begin{equation}
    \Phi e^{2K}\rho=h(t\mp x),
    \label{cons2}
\end{equation}
 in terms of a free function $h$ of one outgoing (ingoing) lightlike coordinate.

 Finally, the set of Einstein equations for pure radiation can be
 cast in the form
 \begin{subequations}
 \begin{eqnarray}
 &&
 U_{tt}-U_{xx}+\frac{1}{\rho}(U_{t}\rho_{t}-U_{x}\rho_{x})=0,\label{U}
 \\
 && \rho_{tt}-\rho_{xx}=0,\label{rho}
 \\
 &&
 K_{t}\rho_{x}+K_{x}\rho_{t}=\rho_{tx}+U_{t}\rho_{x}+U_{x}\rho_{t}+2
 \rho U_{t}U_{x} \mp h,\label{Kt}
 \\
 &&
 K_{t}\rho_{t}+K_{x}\rho_{x}=\frac{\rho_{tt}+\rho_{xx}}{2}+
 U_{t}\rho_{t}+U_{x}\rho_{x}
 +\rho\left(U_{t}^2+U_{x}^2\right)+h,\label{Kx}
 \\
 &&
 K_{xx}-K_{tt}+\frac{U_{x}\rho_{x}-U_{t}\rho_{t}}{\rho}+U_{x}^2-U_{t}^2=0, \label{nu}
 \end{eqnarray}
 \end{subequations}
where the integrability condition for (\ref{Kt}-\ref{Kx}) is the Euler
equation (\ref{cons1}) and the 1-D inhomogeneous wave equation (\ref{nu})
for $K$ is a consequence of the rest of the differential system and
therefore can be dropped. The remaining set of differential equations for 
pure radiation
models is then reduced to a 2-D wave equation for $U$, a 1-D wave equation for
$\rho$ and a quadrature defining $K$.

\section{Simplification of the equations}

The system of differential equations may be simplified by taking
$\rho$ as a coordinate for the equations. Three possibilities arise
according to the character of the gradient of $\rho$:

\begin{itemize}
    \item  If the gradient is lightlike, $\rho_{t}=\varepsilon
    \rho_{x}$, $\varepsilon^2= 1$, the 1-D wave equation is
    trivially satisfied, $\rho_{xx}=\varepsilon \rho_{tx}=\varepsilon
    \rho_{xt}=\rho_{tt}$. We have then $\rho=\rho(t+\varepsilon x)$
    and the equations defining the quadrature,
    \begin{subequations}
    \begin{eqnarray}
      (K_{t}+\varepsilon K_{x})\frac{\rho_{t}}{\rho}&=&
        \frac{\rho_{tt}}{\rho}+(U_{t}+\varepsilon
        U_{x}) \frac{\rho_{t}}{\rho}+2\varepsilon U_{t}U_{x}\mp \varepsilon\Phi e^{2K},\\
        \label{quad1}
    (K_{t}+\varepsilon K_{x})\frac{\rho_{t}}{\rho}&=&
        \frac{\rho_{tt}}{\rho}+(U_{t}+\varepsilon
        U_{x}) \frac{\rho_{t}}{\rho}+U_{t}^2+U_{x}^2+ \Phi e^{2K},
        \label{quad2}
    \end{eqnarray}
    \end{subequations}
allow us to draw rather constraining consequences,
\begin{subequations}
\begin{eqnarray}
    &&\Phi e^{2K} (1\pm\varepsilon)+(U_{t}-\varepsilon U_{x})^2=0,\\&&
    K_{t}\mp K_{x}=
        \frac{\rho_{tt}}{\rho_{t}}+2U_{t}
        +2\rho \frac{U_{t}^2}{\rho_{t}}+ \rho\frac{\Phi e^{2K}}{\rho_{t}},
\end{eqnarray}
\end{subequations}
since $\varepsilon=\mp 1$ in order to have $\Phi\neq0$ and the
gradient of $U$ is also lightlike, $U_{t}=\mp U_{x}$, and parallel to
the gradient of $\rho$, but not to the velocity of the radiation.

No further restrictions are implied for $U$, since it satisfies
$U_{tt}=U_{xx}$ and the remaining condition,
\begin{equation}
    0=\rho_{x}U_{x}-\rho_{t}U_{t}=\langle
    \mathrm{grad}U,\mathrm{grad}\rho\rangle
\end{equation} is trivial.

For $K$ the 1-D wave equation (\ref{nu}) is no longer inhomogeneous,
\begin{equation}
    K_{tt}-K_{xx}=0,
\end{equation}
and we may write the solution in terms of ingoing and outgoing
lightlike coordinates, $u=t\mp x$, $v=t\pm x$,
\begin{equation}
    K(u,v)=f(u)+g(v),
\end{equation}
but just $f$ is constrained by the quadrature,

\begin{equation}
    f'=\frac{\rho''}{2\rho'}+U'+\frac{\rho U'^2}{\rho'}+\frac{h}{2\rho'},
    \label{f}
\end{equation} where the prima denotes derivation with respect to $u$.

In principle, we have four independent functions, $f$, $g$, $\rho$
and $U$ and the system is solved by prescribing

\begin{equation}
    h=2\rho'(f'-U')-\rho''-2\rho U'^2.
    \label{h}
\end{equation}

The metric can be therefore written as

\begin{equation}
    ds^2=-e^{2f+2g}du dv+\rho^2e^{2U}dy^2+e^{-2U}dz^2,
\end{equation}
in terms of lightlike coordinates.

But we see that $f$ and $g$ appear due to just coordinate choices. We may
eliminate them by respectively redefining $u$ and $v$ as $\int du\, e^{2f}$ and
$\int dv\, e^{2g}$ and writing the metric,

\begin{equation}\label{lightmetric}
    ds^2=-2du dv+F^2dy^2+G^2dz^2,
\end{equation}
in terms of just two independent functions, $F$ and $G$, of $u$.

In this simpler frame the function $\Phi$,
\begin{equation}\label{phiphi}
    \Phi=-\left(\frac{F''}{F}+\frac{G''}{G}\right),
\end{equation}
just imposes the positiveness requirement.

It is clear that these spacetimes possess more isometries than those
which were required from the beginning. Besides the obvious Killing
fields, $\{\partial_{v}, \partial_{y}, \partial_{z}\}$ there are
two additional independent isometry generators, $\{y\partial_{v}+\int
F^{-2}\,\partial_{y}, z\partial_{v}+\int G^{-2}\,\partial_{z}\}$ and
therefore the isometry group is a $G_{5}$ acting on null orbits. The spacetimes are not
however homogeneous, since the action of the group of isometries is not
transitive. The Petrov type is generically I.

The Killing field $\partial_{v}$ is covariantly constant and
therefore these models are plane-fronted gravitational waves with
parallel rays spacetimes \cite{brinkmann}, \cite{baldwin}, \cite{kramer}. There are examples where the isometry group is
even larger, for
instance when $F=G$, there is an additional independent Killing
field, $y\partial_{z}-z\partial_{y}$. In this case the spacetime is
conformally flat.

The results may be extended to non-diagonal metrics,

\begin{equation}
    ds^2=-2du dv+F^2dy^2+2 Hdydz+G^2dz^2,
\end{equation}
by just introducing another function, $H$, of $u$.
 
\item  If the gradient of $\rho$ is spacelike, we may choose $\rho$
as a coordinate, $\rho=x$. Equation (\ref{rho}) allows us to keep the
isothermal parametrization. The system thereby becomes much simpler,

\begin{subequations}
\begin{eqnarray}
&&
U_{tt}-U_{xx}-\frac{U_{x}}{x}=0,\label{U2}
\\
&&
K_{t}=U_{t}+2x U_{t}U_{x} \mp h,\label{Kt2}
\\
&&
K_{x}=U_{x}+x\left(U_{t}^2+U_{x}^2\right)+ h,\label{Kx2}
\end{eqnarray}
\end{subequations}
which tell us that radiation fields can be generated from vaccuum
spacetimes by adding a term $\mp H(t\mp x)$ to the conformal factor
$K$ \cite{rao}.

This is the only case where a true axis might appear at $x=0$. But on
applying the flatness condition in the vicinity of the axis
\cite{kramer},

\begin{equation}
    \lim_{x\to0}\frac{\langle
    \textrm{grad}\,\Delta,\textrm{grad}\,\Delta\rangle}{4\Delta}=
    e^{2(U-K)}|_{x=0}=1,
    \quad
    \Delta=\langle\partial_{x},\partial_{x}\rangle=x^2e^{2U},
    \label{axis}
\end{equation}
assuming that the metric functions are regular at $x=0$, the
quadrature equations for $K$, (\ref{Kt2}-\ref{Kx2}), would imply

\begin{equation}
    K_{t}(t,0)=U_{t}(t,0) \mp H_{t}(t),\qquad K_{x}(t,0)=U_{x}(t,0) \mp
    H_{x}(t)
\end{equation}
naming $h=H'$. Therefore $K(t,0)=U(t,0)+H(t)$
and condition (\ref{axis}) cannot be satisfied for all times unless
$H$ is constant.

Furthermore, this means a problem in our parametrization, since the
function $h(t\mp x)$ must be positive (negative) for positive (negative)
values of $x$. Therefore, if $h(t_{0}\mp x_{0})$ is positive for some
values $x_{0}$, $t_{0}$, it should be positive for all values of $t$ and
$h$ would be everywhere positive. It is obvious that there is no place
at the same time for positive and negative values of $x$ and we have
to resort, although there is no axis in strict sense, to the axial
interpretation of the models and admit that $x$ ranges from zero to $\infty$
and it is a sort of radius. The coordinate $y$ would be interpreted as
as an angle. In such case, $H_{t}=H'=h$ must be positive everywhere
and therefore the conformal factor $H$ is a growing function of its
only variable.

    \item  Similarly, if the gradient of $\rho$ is timelike, we may choose $\rho$
as a coordinate, $\rho=t$ without losing generality,

\begin{subequations}
\begin{eqnarray}
&&
U_{tt}-U_{xx}+\frac{U_{t}}{t}=0,\label{U3}
\\
&&
K_{x}=U_{x}+2t U_{t}U_{x} \mp h,\label{Kx3}
\\
&&
K_{t}=U_{t}+t\left(U_{t}^2+U_{x}^2\right)+ h,\label{Kt3}
\end{eqnarray}
\end{subequations}
for which the results of the previous case are also valid.

\end{itemize}

\section{Geodesic completeness of the models}

The most usual definition of absence of singularities is causal
geodesic completeness. This means that every timelike and lightlike
geodesic, that is every physical trajectory of a test particle in the spacetime
in the absence of external non-gravitatory forces, can be extended
from $-\infty$ to $\infty$ in their affine parametrization,
$x^{\mu}=x^\mu(\tau)$,

\begin{eqnarray}
    &&\ddot x^\mu+\Gamma^\mu_{\nu\rho}\dot x^\nu \dot x^\rho=0, \quad
     g_{\mu\nu}\dot x^\nu \dot x^\rho=-\delta,\quad
    -\infty<\tau<\infty
\end{eqnarray}
where $\delta$ takes the value zero for lightlike and one for timelike
geodesics and the dot stands for derivation with respect to the affine
parameter $\tau$. It is called affine parameter because it is unique
but for affine transformations.

Causal geodesic completeness implies a simpler definition, regularity
of the curvature scalars, but it is not true the other way round.
Taub-NUT spacetime \cite{nut} is a nice counterexample for this \cite{HE}.
Moreover, spacetimes where curvature scalars are regular may be
incomplete because they can be extended to larger manifolds.
Therefore in order to check geodesic completeness we are to analyse
geodesic equations.

We have already seen that our models are classified in three sets
according to the character of the gradient of $\rho$:

\begin{itemize}
    \item  Lightlike gradient of $\rho$: In this case the analysis of
    the geodesics becomes quite simple, since there are extra
    isometries, which allow us to reduce the order of the geodesic
    equations. We just need the three commuting generators
    $\{\partial_{v}, \partial_{y}, \partial_{z}\}$ to derive
    independent constants of geodesic motion,

    \begin{eqnarray}
    V & = & \langle \partial_{v},\dot x\rangle=
    g_{v\mu}\dot x^\mu= -\dot u,\\
        Y & = & \langle \partial_{y},\dot x\rangle=
        g_{y\mu}\dot x^\mu= F^2\dot y,  \\
    Z & = & \langle \partial_{z},\dot x\rangle=
    g_{z\mu}\dot x^\mu= G^2\dot z,
    \end{eqnarray}

    The first  equation simply implies that $u$ is an acceptable
    affine parameter for the geodesics,

    \begin{equation}
        u=u_{0}-V\,\tau.
    \end{equation}

    The second equation can be integrated for all values of $\tau$ or
    $u$,

    \begin{equation}
    y=y_{0}-\frac{Y}{V}\int du\,F^{-2}(u).
    \end{equation}
    provided that $F^{-2}$ is integrable for all values of $u$.

    The third equation is also integrable if  $G^{-2}$ is integrable
    for all values of $u$,

    \begin{equation}
    z=z_{0}-\frac{Z}{V}\int du\,G^{-2}(u).
    \end{equation}

    The remaining equation of geodesic motion needs not be derived
    from the Christoffel symbols but from the condition of affine
    parametrization,

    \begin{equation}
        \delta =- 2V\dot v-Y^2F^{-2}-Z^2G^{-2},
    \end{equation}
    from which we do not get any additional restriction,
    \begin{equation}
    \frac{dv}{du} = \frac{1}{2V^2}\left\{\delta+Y^2F^{-2}+Z^2G^{-2}\right\}.
    \end{equation}
    since $v$ is integrable provided that $y$ and $z$ are integrable.

    These results may be summarized in the following theorem:

     \begin{theorem}
         An spacetime with a metric of the form
        (\ref{lightmetric}) filled with a pure radiation field is causally
        geodesically complete if and only if the metric functions
        $g^{yy}$ and $g^{zz}$ are integrable for all values of the
        light coordinate $u$.
    \end{theorem}


    Examples of such spacetimes are easy to provide. For instance, if
    we resort to powers of the coordinate $u$, $F(u)=|u|^{\alpha/2}$,
    $G(u)=|u|^{\beta/2}$, the regularity requirement is fulfilled if
    $\alpha, \beta<1$. Since $\Phi$ has to be positive, an additional
    condition is imposed, $0<\alpha, \beta<2$. This example is
    intriguing since it states that half of the models are regular
    whereas the other half are singular.

    Although the previous reasoning has been made for specific 
    functions, it describes the behaviour of general functions $F$ 
    and $G$ on becoming zero. In fact, since equation (\ref{phiphi}) 
    imposes that the graphic of at least one of the functions must be 
    convex, it is clear that the graphic of this function is to cut the axis at one 
    point at least.
    
    This poses a serious drawback on these spacetimes, since the 
    previous analysis points out that the metric functions $F^2$ and 
    $G^2$ are continuous, but not differentiable. We are dealing with 
    topological manifolds instead of differentiable manifolds. 
    However, metric invariants do not see this fact, since they are 
    all null.

    \item  The analysis of models with non-lightlike gradient of
    $\rho$ is quite more complicated, since generically we do not
    have additional symmetries. Fortunately, we may resort to some
    theorems which have been explicitly devised for $G_{2}$
    spacetimes \cite{manolo};


    \begin{description}
    \item[Theorem:] A diagonal Abelian orthogonally transitive spacetime
    with spacelike orbits endowed with a metric in the form (\ref{metric}) with $C^2$
    metric functions  $K,U,\rho$, where $\rho$ has a spacelike gradient, is future causally geodesically complete
    provided that along causal geodesics:
    \begin{enumerate}
    \item For large values of $t$ and increasing $x$,
    \begin{enumerate}
    \item $(K-U-\ln\rho)_{t}+(K-U-\ln\rho)_{x}\ge 0$, and either
    $(K-U-\ln\rho)_{x}\ge 0$  or $|(K-U-\ln\rho)_{x}|\lesssim
    (K-U-\ln\rho)_{t}+(K-U-\ln\rho)_{x}$.
    \item $K_{t}+K_{x}\ge 0$, and either $K_{x}\ge 0$ or
    $|K_{x}|\lesssim K_{t}+K_{x}$.
    \item $(K+U)_{t}+(K+U)_{x}\ge 0$, and either $(K+U)_{x}\ge 0$
    or $|(K+U)_{x}|\lesssim (K+U)_{t}+(K+U)_{x}$.
    \end{enumerate}


    \item \label{tt} For large values of  $t$, a constant $b$ exists such that
    \\
    $\left.\begin{array}{c}K(t,x)-U(t,x)\\2\,K(t,x)\\K(t,x)+U(t,x)+\ln\rho(t,x)
    \end{array}\right\}\ge-\ln|t|+b.$

    \end{enumerate}\end{description}

    For geodesics pointing to the past the theorem imposes the same
    conditions but for a sign in the time derivatives.

    We just have to check the restrictions on the metric functions
    that such conditions require:

    \begin{enumerate}
    \item Since $h$ is positive for radiation fields and positive
    values of $x$, we may reach some conclusions:
    \begin{enumerate}
    \item
    $(K-U-\ln\rho)_{t}+(K-U-\ln\rho)_{x}\ge
    x\left(U_{t}+U_{x}\right)^2-{1}/{x}$ is positive
    unless the derivatives of $U$ decrease too quickly. Such possibility
    is avoided if we impose

    \begin{equation}
        x^{1-\varepsilon}|U_{x}+U_{t}| \not\to 0,
        \label{cond1}
    \end{equation}
    for large values of $t$ and $x$.

    On the other hand $(K-U-\ln\rho)_{x}\ge
    x\left(U_{t}^2+U_{x}^2\right)-{1}/{x}$ is also positive
    under the same conditions.

    \item $K_{t}+K_{x}\ge U_{t}+U_{x}+x\left(U_{t}+U_{x}\right)^2$
    is again positive for large values of $t$ and $x$ if (\ref{cond1}) is
    satisfied, as it happens with $K_{x}\ge U_{x}+x\left(U_{t}^2+U_{x}^2\right)$.

    \item The same restrictions are valid for $(K+U)_{t}+(K+U)_{x}$ and
    $|(K+U)_{x}|$. Therefore (\ref{cond1}) suffices for the satisfaction
    of the first part of the theorem.
    \end{enumerate}

    For past-pointing geodesics, as it has already been told,
    condition 1 is the same for large negative values of $t$, but for the sign of the time derivatives.
    Therefore, the corresponding requirement for past-pointing
    geodesics is

    \begin{equation}
    x^{1-\varepsilon}|U_{x}-U_{t}| \not\to 0,
        \label{cond11}
    \end{equation}
    for small values of $t$ and large values of $x$.

%
%
%
%

    \item On $x=0$ we already know that $K(t,0)-U(t,0)=\mp
    H(t)$ and $H$ is a growing function. Therefore we
    may encounter a problem for outgoing radiation unless $H$ grows
    slower than a logarithm for large times,
    \begin{equation}
    \mp  H(t)\ge-\ln|t|+b.\label{hres}
    \end{equation}

    Out of $x=0$ the first
    condition of this set is satisfied,
 \begin{equation}
     K(t,x)-U(t,x)=\mp H(t\mp x)+\int_{0}^x x'\left\{U_{t}(t,x)^2+
     U_{x}(t,x)^2\right\}.
 \end{equation}
since we are adding just a positive term to the previous case.

    For the second condition on $x=0$, we obtain another constraint
    for large values of $t$,

    \begin{equation}
    K(t,0)=U(t,0)\mp  H(t)\ge-\frac{1}{2}\ln|t|+b,
    \end{equation}
   and out of $x=0$ it will be also satisfied.

   Finally, for the third condition we may require,
   \begin{equation}
       2U(t,0)\mp  H(t)\ge-\ln|t|+b,
   \end{equation}
   always for large values of $t$.

    \end{enumerate}

These conditions on outgoing radiation may be summarized in the following
way for large values of $t$,

\begin{eqnarray}
    H(t) & \le & \ln|t|+b \nonumber\\
    U(t,0) & \ge & b-\frac{1}{2}\ln|t|+ H(t),
    \label{out}
\end{eqnarray}
 since $H$ grows obviously faster than $H/2$.

For ingoing radiation condition (\ref{hres}) is always fulfilled,
since $H$ is a growing function. The only requirement which remains is

\begin{eqnarray}
    U(t,0)  \ge  b-\frac{1}{2}\ln|t|-\frac{H(t)}{2},
    \label{in}
\end{eqnarray}
for large values of $t$.

Similar restrictions are imposed for past-pointing geodesics for large
negative values of $t$. The roles of outgoing and ingoing radiation
are obviously exchanged.

For outgoing radiation the only restriction in order to achieve
causal geodesic completeness to the past is

\begin{eqnarray}
    U(t,0)  \ge  b-\frac{1}{2}\ln|t|+\frac{H(t)}{2},
    \label{in2}
\end{eqnarray}
for small values of $t$, since $H/2$ decreases to the past more slowly than $H$.

On the other hand, the corresponding condition for ingoing radiation is

\begin{eqnarray}
    H(t) & \ge & -\ln|t|+b \nonumber\\
    U(t,0) & \ge & b-\frac{1}{2}\ln|t|-H(t),
    \label{out2}
\end{eqnarray}
for small values of $t$.

Summarizing the previous results, we may enclose them in two statements:

\begin{theorem}
      A diagonal Abelian orthogonally transitive spacetime
    with spacelike orbits with a metric in the form (\ref{metric})
    and a spatial gradient of $\rho$ and a pure
    outgoing radiation field as matter content is causally geodesically
    complete if:

    \begin{enumerate}
        \item  $x^{1-\varepsilon}|U_{x}\pm U_{t}| \not\to 0$ for large values
    of $|t|$ and $x$.

        \item  $H(t) \le \ln|t|+b$ and $U(t,0) \ge  b-\frac{1}{2}\ln|t|+
        H(t)$ for large values of $t$.

    \item $U(t,0) \ge  b-\frac{1}{2}\ln|t|+ H(t)/2$ for small values of $t$.
    \end{enumerate}
\end{theorem}

\begin{theorem}
      A diagonal Abelian orthogonally transitive spacetime
    with spacelike orbits with a metric in the form (\ref{metric})
    and a spatial gradient of $\rho$ and a pure
    ingoing radiation field as matter content is causally geodesically
    complete if:

    \begin{enumerate}
    \item  $x^{1-\varepsilon}|U_{x}\pm U_{t}| \not\to 0$ for large values
    of $|t|$ and $x$.

    \item  $U(t,0)  \ge  b-\frac{1}{2}\ln|t|-H(t)/2$ for large values of $t$.

    \item $H(t) \ge -\ln|t|+b$ and $U(t,0) \ge b-\frac{1}{2}\ln|t|-H(t)$ for small values of $t$.
    \end{enumerate}
\end{theorem}

The meaning of such results seems clear. There must be an upper limit
for $H$ in the future for outgoing radiation in order to prevent a
huge amount of energy that could form a singularity. A symmetric
reasoning can be applied to ingoing radiation.

    $H$ can be any growing function, but $U$ is restricted by the 2-D
    wave equation (\ref{U2}). Following \cite{john}, we may write the general
    solution of this equation in terms of initial data,
    $U(0,x)=f(x)$, $U_{t}(0,x)=g(x)$,

    \begin{equation}
    U(t,x)=\frac{1}{2\pi}\int_{0}^{2\pi}d\phi\int_0^1d\tau
     \frac{\tau}{\sqrt{1-\tau^2}}\left\{tg(v)+f(v)+tf'(v)
     \frac{t\tau^2+x\tau\cos\phi }{v}
    \right\},\end{equation}
    where $v=\sqrt{x^2+t^2\tau^2+2xt\tau\cos\phi}$.

    Since we are interested in the values of $U(t,0)$, the expression
    we need is much simpler,
    \begin{equation}
    U(t,0)=\int_0^1d\tau
     \frac{\tau}{\sqrt{1-\tau^2}}\left\{tg(|t|\tau)+f(|t|\tau)+|t|\tau
     f'(|t|\tau)
    \right\},\end{equation}
    which allows easy implementation of the regularity conditions.

    For instance, we can construct non-singular radiation models just
    prescribing a function $U$ which grows for large values of $|t|$
    and $x$ and a function $H$ that does not grow or decrease too
    fast, that is, bounded by a logarithm.

    A simple example of a way to comply with the regularity
    requirements for $U$ is shown in \cite{wide} for polynomial
    initial data: if $f(x)=\sum a_{i}x^i$ and $g(x)=\sum b_{j}x^j$ are polynomials in $x$ of degrees
    $m$, $n$, respectively, $U$ fulfills the regularity conditions
    if $a_{m}>0$ and $m>n+1$ or if $m=n+1$ and $(m+1/2)a_{m}>|b_{n}|$.

    \item  We shall not consider causal geodesic completeness in the
    case of a timelike gradient of $\rho$ since the logarithm term of
    the time coordinate in the regularity theorems would mean a
    singularity at $t=0$.
\end{itemize}

\section{Discussion}

We have discussed in this paper the possibility of constructing
regular inhomogenous spacetimes with a pure radiation field as matter
content. It has been shown that such cosmological models are very easy
to construct in terms of a nearly arbitrary growing function $H$,
which is bounded by a logarithmic function, and a solution of a wave
equation, $U$, with rather few limitations. The conclusions are close
to the results for stiff perfect fluids: pure radiation inhomogeneous
cosmological models are an infinite family and are not negligible in
the set of much models.

Pure radiation fields satisfy all energy conditions and these models
are causally stable since the function $t$ has a positive gradient in
all the spacetime \cite{HE}. Therefore these models fulfill weaker causality
conditions suchs as the chronology conditions. As it happened in most
regular cosmological models, the way to elude singularity theorems is
the lack of trapped sets.

The existence of such an amount of non-singular cosmological models
fulfilling all energy and chronology conditions
suggests that they are more abundant than it was thought before. If
they had not been found previously it was due merely to the techniques
used to generate them, not to their scarcity among inhomogeneous
cosmological models. It remains however an open question whether such
non-singular behaviour is just a peculiarity of $G_{2}$ inhomogeneous
models and if these results can be extended to smaller groups of
isometries or to other matter contents.

As a curiosity, it is interesting to notice that perfectly regular spacetimes can be
turned singular by introducing radiation. For instance, the metric,

\begin{equation}
    ds^2=e^{r-t}\left\{-dt^2+dr^2\right\}+r^2d\phi^2+dz^2,
 \end{equation}
 can be generated from Minkowski spacetime by adding a conformal
 factor. It is easy to check that axial geodesics are not complete,

 \begin{equation}
     r=0,\quad \dot\phi=0, \quad \delta=e^{-t}\dot t^2-Z^2,
  \end{equation}
  \begin{equation}
      \dot t=\sqrt{\delta+Z^2}e^t\Rightarrow t(\tau)=-
      \ln\left(\sqrt{\delta+Z^2}(k-\tau)\right).
   \end{equation}

\section*{Acknowledgments}
The present work has been supported by Direcci\'on General de
Ense\~nanza Superior Project PB98-0772. The authors wish to thank
 F.J. Chinea and  F. Navarro-L\'erida for valuable discussions.

\end{document}